%% file: main.tex
\documentclass[journal]{IEEEtran}
\input{include/0_Packages.tex}
\input{MySettings.tex}

\begin{document}
%
\title{On-Chip Characterization of High-Loss Liquids between 750~GHz and 1100~GHz}

%
%
%

\newcommand{\orcidauthorA}{0000-0002-9821-3199} 
\newcommand{\orcidauthorB}{0000-0002-5861-6655} 
\newcommand{\orcidauthorC}{0000-0001-9519-3527} 
\newcommand{\orcidauthorD}{0000-0002-8204-7894} 

\author{Juan~Cabello-S\'{a}nchez,~\IEEEmembership{Student Member,~IEEE,}
Vladimir~Drakinskiy,
Jan~Stake,~\IEEEmembership{Senior~Member,~IEEE,}
Helena~Rodilla,~\IEEEmembership{Senior~Member,~IEEE}
\thanks{Manuscript received July 17, 2020; revised September 5, 2020; accepted October 1, 2020. This work was supported by the Swedish Research Council (Vetenskapsr\aa det) under grant 2015-03981
, and Knut and Alice Wallenberg Foundation under grant 2014.0275.}
\thanks{Juan Cabello-S\'{a}nchez, Vladimir Drakinskiy, Jan Stake and Helena Rodilla are with the Terahertz and Millimetre Wave Laboratory, Chalmers University of Technology, SE-412 96 Gothenburg, Sweden. (e-mail: \mbox{juancab@chalmers.se}; \mbox{vladimir.drakinskiy@chalmers.se}; \mbox{jan.stake@chalmers.se}; \mbox{rodilla@chalmers.se})}
\thanks{Color versions of one or more of the figures in this article are available online at http://ieeexplore.ieee.org.}
\thanks{Digital Object Identifier 10.1109/TTHZ.2020.3029503}}%
%
%

\markboth{IEEE Transactions on Terahertz Science and Technology,~Vol.~11 No.~1, January~2021}%
{Cabello-S\'{a}nchez \MakeLowercase{\textit{et al.}}: On-Chip Characterization of High-Loss Liquids between 750~GHz and 1100~GHz}
%



\maketitle

\input{include/1_Abstract.tex}

%
\IEEEpeerreviewmaketitle

\input{include/2_Introduction.tex}

\input{include/3_Method.tex}

\input{include/4_Results.tex}

\input{include/5_Conclusions.tex}



%

\appendices
\input{include/6_Appendices.tex}

\input{include/7_Acknowledment.tex}

\ifCLASSOPTIONcaptionsoff
  \newpage
\fi



%


\bibliographystyle{IEEEtran} 
\bibliography{bibtex/bib/references_new.bib}

\input{include/8_Biography.tex}

\end{document}

%% file: include/0_Packages.tex
\usepackage{cite}

\ifCLASSINFOpdf
\else
\fi
\usepackage{amsmath}

%% file: MySettings.tex
\usepackage{amsfonts} 
\usepackage{color}
\usepackage[version=4]{mhchem}
\usepackage{pgfplots}
\usepackage{siunitx}
\usepackage{circuitikz} 
    \usetikzlibrary{decorations.markings}
    \usepackage{adjustbox}
    \usepackage{adjustbox}
\usepackage{hyperref}
\usepackage{stix}

%% file: include/1_Abstract.tex
\begin{abstract}
Terahertz spectroscopy is a promising tool for analyzing the picosecond dynamics of biomolecules, which is influenced by surrounding water molecules.
However, water causes extreme losses to terahertz signals, preventing sensitive measurements at this frequency range.
Here, we present sensitive on-chip terahertz spectroscopy of highly lossy aqueous solutions using a vector network analyzer, contact probes, and a coplanar waveguide with a 0.1~mm wide microfluidic channel.
The complex permittivities of various deionized water/isopropyl alcohol concentration are extracted from a known reference measurement across the frequency range 750--1100~GHz and agrees well with literature data.
The results prove the presented method as a high-sensitive approach for on-chip terahertz spectroscopy of high-loss liquids, capable of resolving the permittivity of water.
\end{abstract}

\begin{IEEEkeywords}
Coplanar waveguides (CPW), isopropyl alcohol (IPA), material properties, microfluidic channels, on-wafer measurements, permittivity, scattering parameters, terahertz spectroscopy, vector network analyzers (VNA), water
\end{IEEEkeywords}

%% file: include/2_Introduction.tex
\section{Introduction}
%
%
%
%
\IEEEPARstart{T}erahertz (THz) spectroscopy is an indispensable tool to analyze light-weight molecules with applications in astronomy and chemistry.
With new technological developments, the application of THz technology has extended to fields as diverse as security \cite{Appleby2007}, communications \cite{Song2011}, pharmaceutical control \cite{Zeitler2007a}, medicine and biology \cite{Pickwell2006}.
In biology, THz waves have shown to be a relevant method for studying picosecond dynamics of biomolecules \cite{Acbas2014, Lundholm2015}, predicted to be key for their biological function \cite{Henzler-Wildman2007} in which water plays an important role~\cite{Xu2015c}.
Aqueous samples have been measured with time-domain spectroscopy (TDS), one of the most common THz spectroscopy methods, with free-space transmission~\cite{Kindt1996},  free-space  reflection~\cite{Thrane1995}, and on-chip setups~\cite{Kitagawa2006, Swithenbank2017}.
However, despite having a relatively high dynamic range, necessary for measuring high-loss samples, the low time-averaged power and wide bandwidth typically yield measurements with a low signal-to-noise ratio (SNR), limiting the smallest detectable signal change \cite{Naftaly2009}.
The low SNR further plummets when measuring high-loss aqueous samples (around 100~dB/mm \cite{Kindt1996}) in higher-loss chip setups, hindering pure water measurements at frequencies above $\SI{0.5}{THz}$ \cite{Swithenbank2017}, or having to avoid liquid sample in the region with the most intense electric field to minimize losses \cite{Ohkubo2006a, Kitagawa2006}, but sacrificing sensitivity.

A promising method for obtaining high SNR for on-chip applications is measuring \mbox{S-parameters}  using vector network analyzers (VNA), a common measuring method at microwave and millimeter-wave frequencies.
This method is based on an electronic heterodyne technique, and benefits from having first-class frequency resolution ($\sim\SI{1}{Hz}$), about \SI{20}{dB} higher dynamic range than TDS systems at \SI{1}{THz} \cite{George2015}, traceability to the International System of Units \cite{Ridler2016}, and can use calibration techniques to move the reference plane to the region of interest \cite{Bauer1974a}; whereas the downside is that the bandwidth is limited to a waveguide band.
The frequency range of VNA measurements applied to biology has been typically restricted to microwave \cite{Velez2017a} and millimeter-wave frequencies  \cite{Rodilla2016a, Bao2018b}.
However, recent development in heterodyne technology has increased the maximum frequency of VNA analysis up to 1.5~THz \cite{Koller2016a} in rectangular waveguides, and up to 1.1~THz for on-chip measurements using contact probes \cite{Bauwens2014}.
Contact probes offer an efficient way to guide the generated power directly into the sensing chip (compared to free-space coupling), thus increasing the sensibility of the method.

In this letter, we demonstrate the use of a vector network analyzer and contact probes for THz spectroscopy of high-loss aqueous samples contained in a chip.
We describe the design of the low-loss sensing waveguide and its fabrication, the measurement setup, and how the complex refractive index was extracted from the measured complex transmission coefficient.
This is a first step towards a miniaturized chip sensor for high-loss liquid samples at THz frequencies.

%% file: include/3_Method.tex
\section{Method}

\begin{figure}[!t]
\centering
\includegraphics[height=0.25\textwidth]{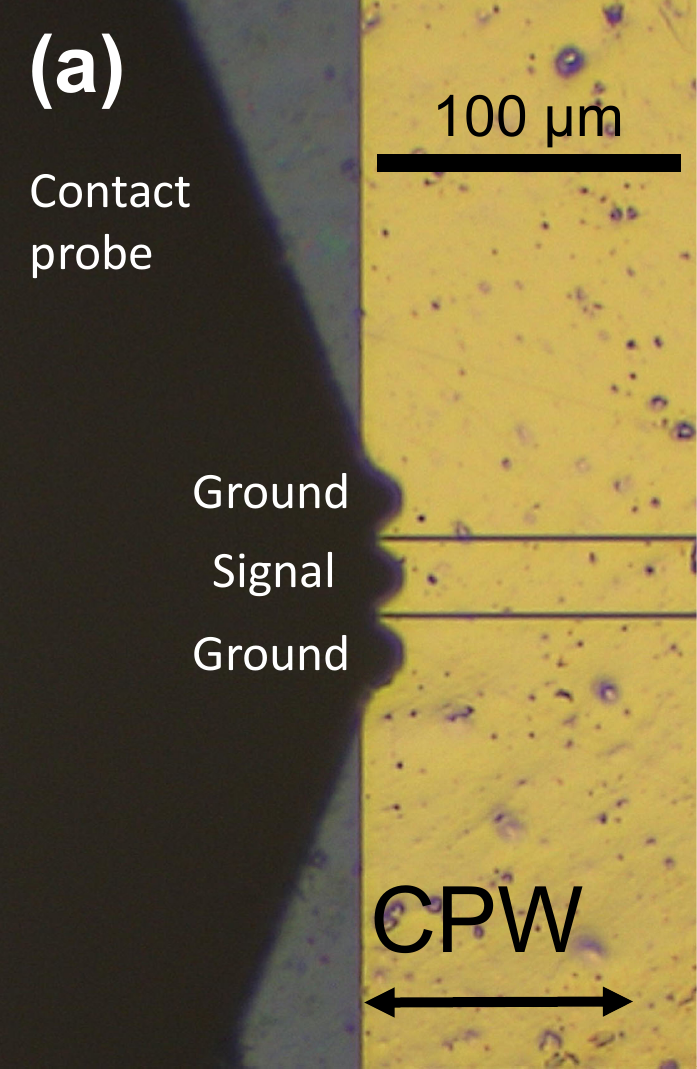}
\includegraphics[height=0.25\textwidth]{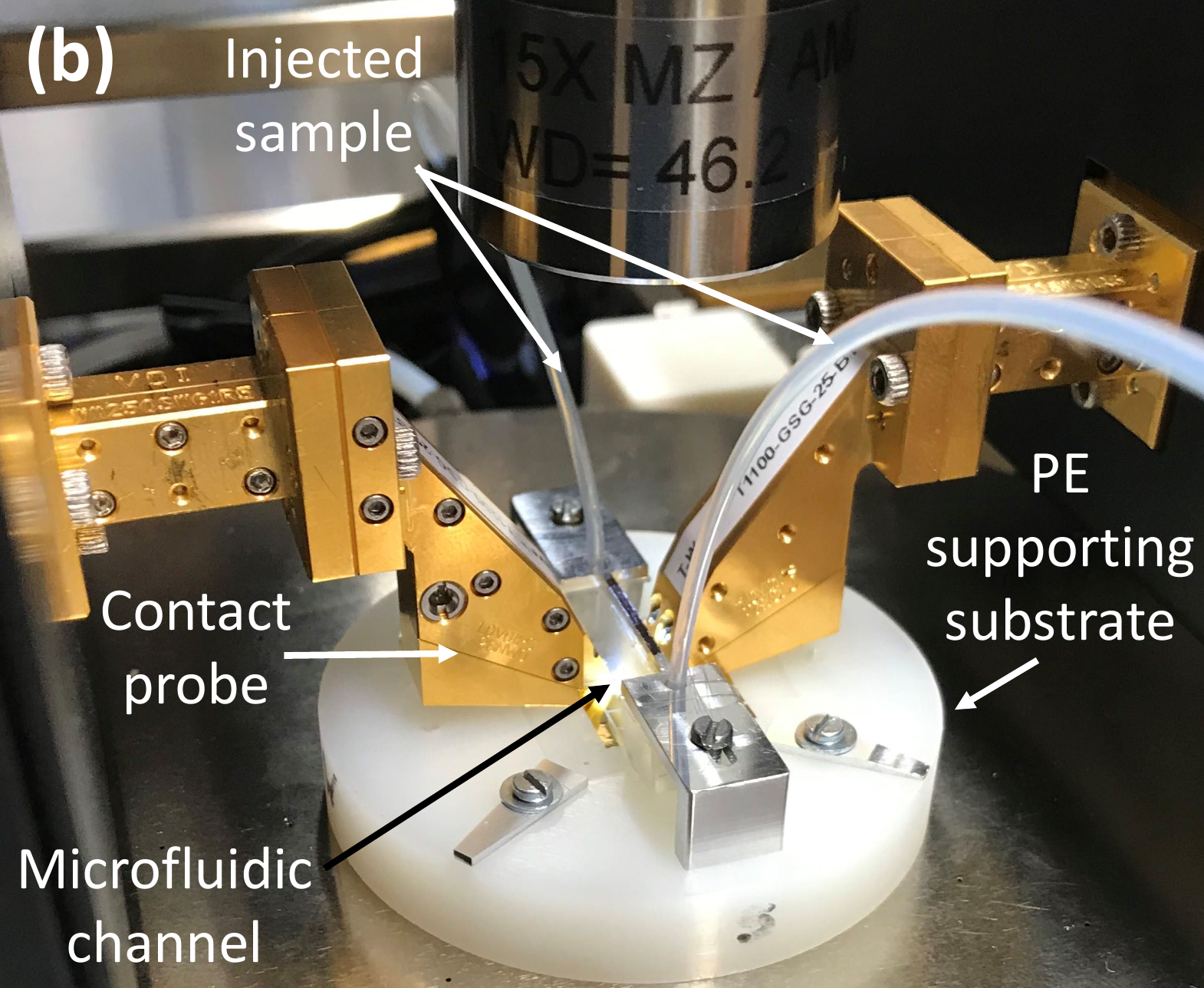}
\caption{(a) Micrograph of the CPW being excited by the ground-signal-ground probe. (b) Photograph of measurement setup during measurements.}
\label{fig:meas_setup_pics}
\end{figure}

For sensing liquid samples on a chip, we used a coplanar waveguide (CPW) \cite{Wen1969}, which provides easy interfacing with ground-signal-ground probes (Fig.~\ref{fig:meas_setup_pics}.a).
The CPW was designed to suit both the measurement probe's pitch ($\SI{25}{\um}$) and characteristic impedance ($\SI{50}{\ohm}$) \cite{Bauwens2014}, yielding a central strip width of $\SI{23.5}{\um}$ and a ground separation of $\SI{1.5}{\um}$.
The CPW was designed on a ultra-thin $\SI{23}{\um}$ thick polyethylene terephthalate (PET) film substrate ($\epsilon'=3.15$ and $\tan(\delta)=0.017$ at 1~THz \cite{Fuse2011}), to avoid power leakage \cite{CabelloSanchez2018} to undesirable substrate modes \cite{Rutledge1978}.
The CPWs and their dedicated calibration standards were fabricated using e-beam lithography and evaporation of $\SI{20}{\nm}$ Ti and $\SI{350}{\nm}$ Au on top of the PET substrate.

\begin{figure}[!t]
\centering
\includegraphics[width=0.48\textwidth]{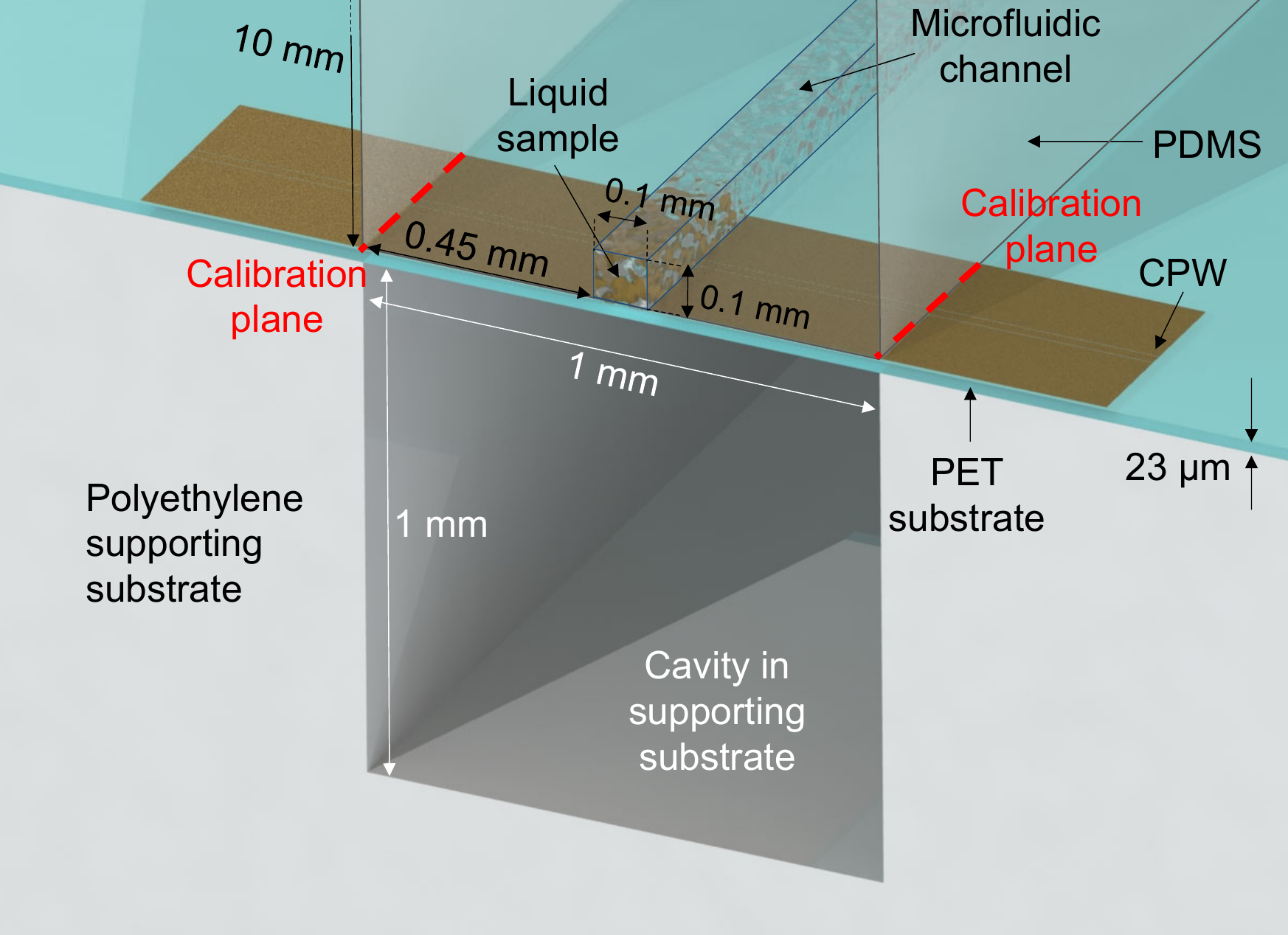}
\caption{Illustration of the cross-section of the microfluidic channel filled with sample intersecting the CPW.}
\label{fig:meas_setup_closeUp}
\end{figure}

The PET substrate containing the CPWs is held on top of a polyethylene supporting substrate to avoid coupling with the probe station's metal chuck during the measurements  (Fig.~\ref{fig:meas_setup_pics}.b).
The polyethylene supporting substrate had a $\SI{1}{\mm}$ deep and $\SI{1}{\mm}$ long air cavity under the measured CPW, suspending it on air (Fig.~\ref{fig:meas_setup_closeUp}).
On top of the PET substrate, an interchangeable polydimethylsiloxane (PDMS) microfluidic channel was clamped to the polyethylene supporting substrate.
The microfluidic channel was designed to have a $\SI{100}{\um}$ wide square cross-section, whereas the cross-section of the PDMS  containing the channel is $\SI{1}{mm}$ wide by $\SI{10}{mm}$ tall (Fig.~\ref{fig:meas_setup_closeUp}).
To deliver the sample into the microfluidic channel, input and output tubings were connected to the microfluidic channel. 

We measured the complex transmission coefficient, $\hat{T}_s$, between  $\SI{0.75}{THz}$ and $\SI{1.1}{THz}$ using a VNA Keysight N5242A connected to two VDI WR1.0SAX frequency extenders \cite{Crowe2011}, having a typical/minimum dynamic range of 65/45~dB, respectively, and a continuous wave signal power higher than $\SI{-40}{dBm}$.
To couple the signal to the sensing chip, we used DMPI T-Wave ground-signal-ground probes \cite{Bauwens2014}.
We used dedicated multi-line TRL \cite{Marks1991b} calibration structures to set the calibration plane at the air-PDMS interface (Fig.~\ref{fig:meas_setup_closeUp}).

The aqueous samples consist of propan-2-ol (IPA) in deionized-water (DI-$\ce{H_2O}$) with different concentrations---0, 10, 20, 30, 40, and 50\% of IPA in volume.
The samples were pumped into the microfluidic channel from higher to lower concentration of IPA through the input tube using a syringe and flushing air between samples.
The pressure was adjusted to atmospheric pressure by removing the syringe momentarily after introducing each sample.
The probes were kept in contact with the chip throughout all measurements to minimize probing position uncertainty.
Each sample was measured five times in consecutive VNA sweeps with an intermediate frequency bandwidth of $\SI{50}{Hz}$.

The effective refractive index of the sample-loaded CPW  ($\hat{n}_{e} = n_{e} -j\kappa_{e}$) was calculated by comparing the measured transmission with a reference measurement, $\hat{T}_r$ (average of five measurements with DI-water as sample), following the equation:
\begin{equation} \label{eq:RI_eff}
\begin{split}
    \hat{n}_{e} &=  \hat{n}_{r} - \frac{\ln{\left[\hat{T_s}/\hat{T_r}\right]}} {jk_0l_s}
\end{split}
\end{equation}
where $k_0$ is the vacuum wavenumber, $l_s$ is the effective sample length, and $\hat{n}_r$ is the effective refractive index of the reference.
$\hat{n}_r$ was obtained from analytical expressions  for multilayered substrates on CPWs \cite{Chen1997} and using a double Debye model for water \cite{Kindt1996} in order to include the frequency-dependent permittivity.

Finally, the sample's refractive index $\hat{n}_s$ was found from measured CPW's effective refractive index, $\hat{n}_e$, by using the same analytical expressions for multilayered substrates on CPWs \cite{Chen1997}. According to it, the equation relating the sample permittivity with the CPW's effective permittivity is:
\begin{equation} \label{eq:RI_smpl}
\begin{split}
    \hat{n}_{e}^2 = \hat{\epsilon}_e = 1 + \frac{1}{2}(\hat{n}_{s}^2-1)\frac{K(k)K(k'_{s})}{K(k')K(k_{s})} + \\ + \frac{1}{2}(\hat{n}_{sub}^2-1)\frac{K(k)K(k'_{sub})}{K(k')K(k_{sub})}
\end{split}
\end{equation}

where $K$ is the complete elliptical integral of the first kind, and $k$, $k'$, $k_{s}$, $k'_{s}$, $k_{sub}$, $k'_{sub}$, are terms depending on the geometry of the CPW cross-section, superstrate (sample) and substrate, respectively \cite{Chen1997}. The analytical expression (eq.~\ref{eq:RI_smpl}) agrees well with more detailed 3D electromagnetic simulations of the multilayered CPW.

%% file: include/4_Results.tex
\section{Results}

\begin{figure}[!t]
\centering
\includegraphics[width=0.49\linewidth, clip=true, trim=6mm 0mm 7mm 0mm]{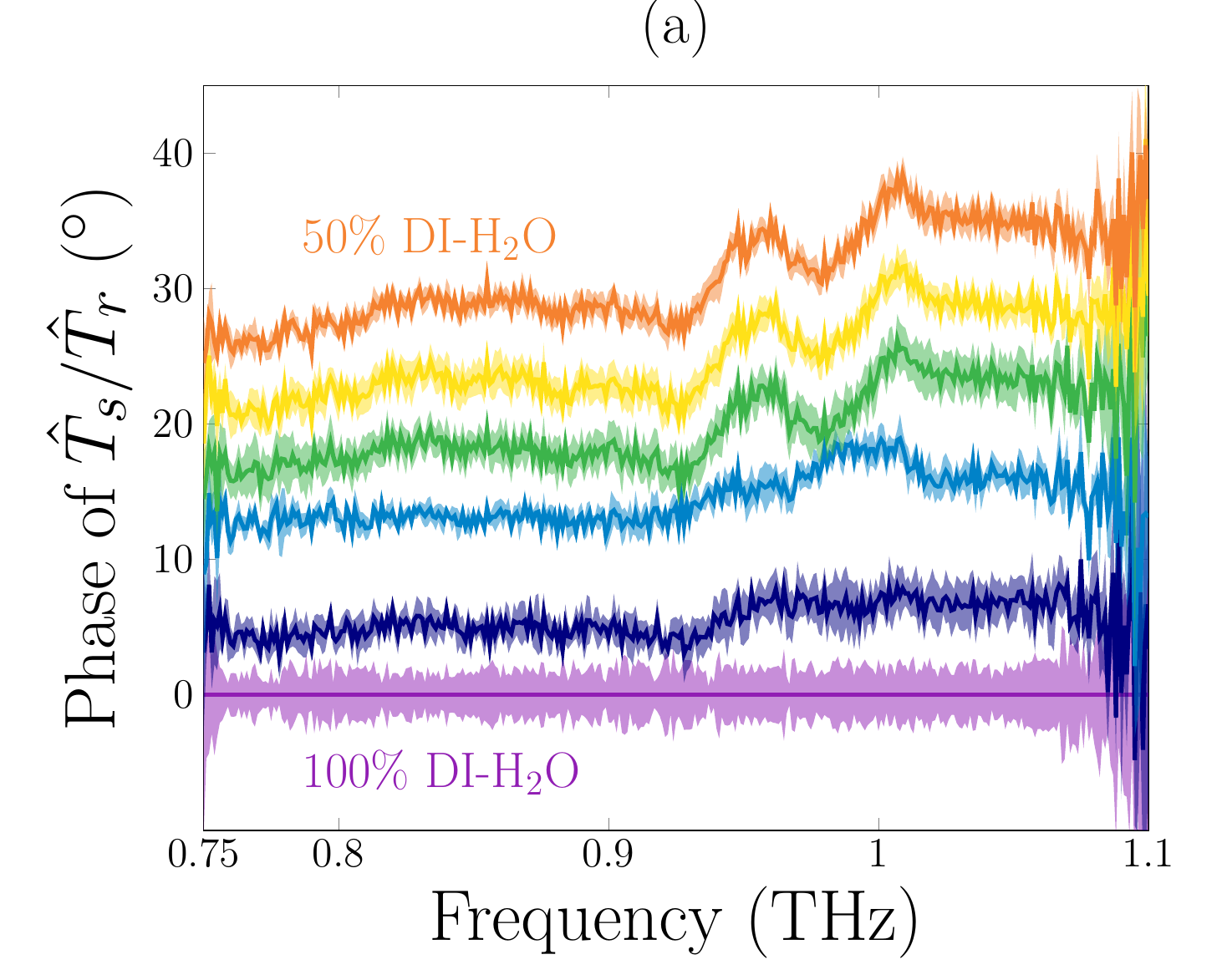}
\includegraphics[width=0.49\linewidth, clip=true, trim=6mm 0mm 7mm 0mm]{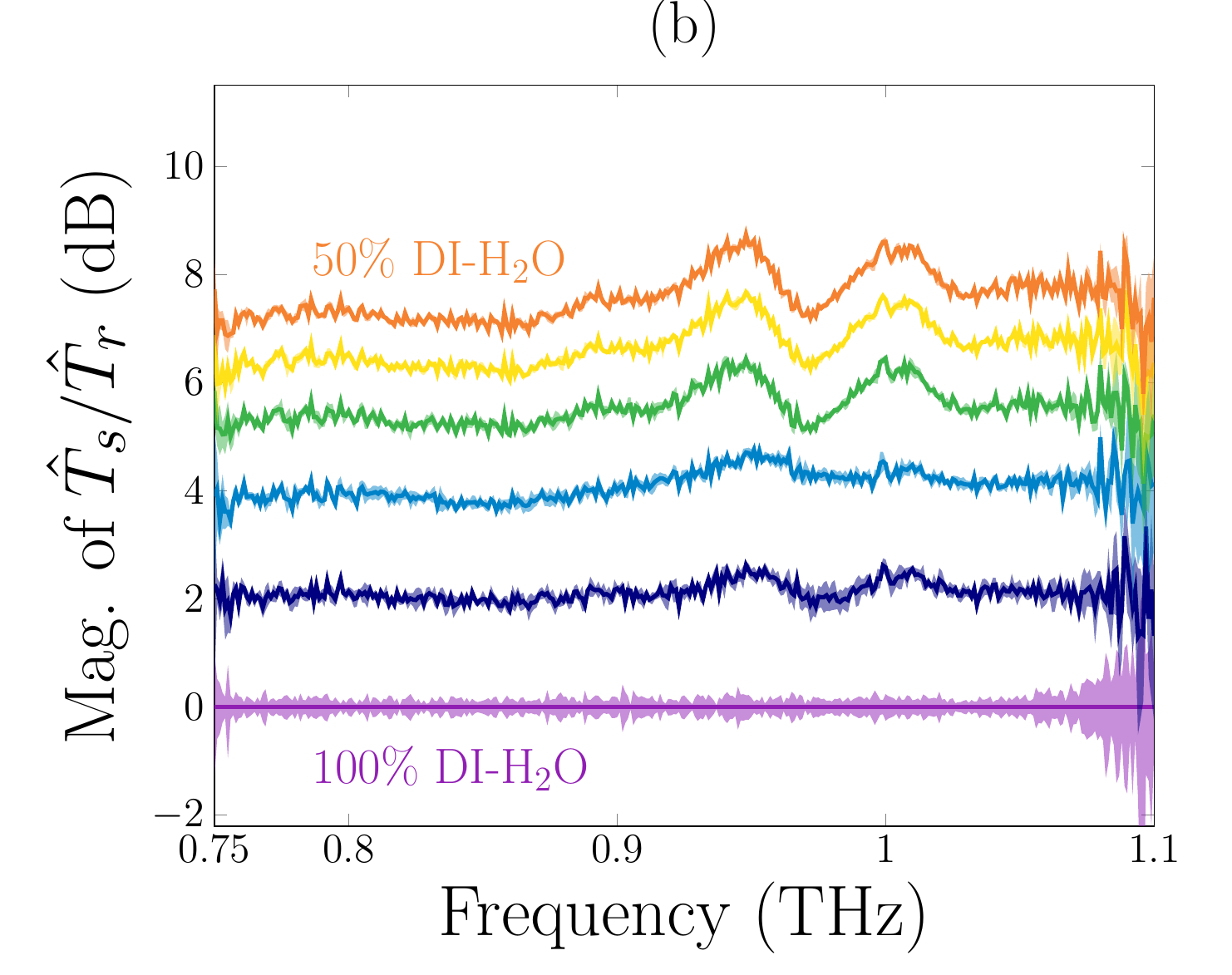}\\
\includegraphics[width=0.49\linewidth, clip=true, trim=6mm 0mm 7mm 0mm]{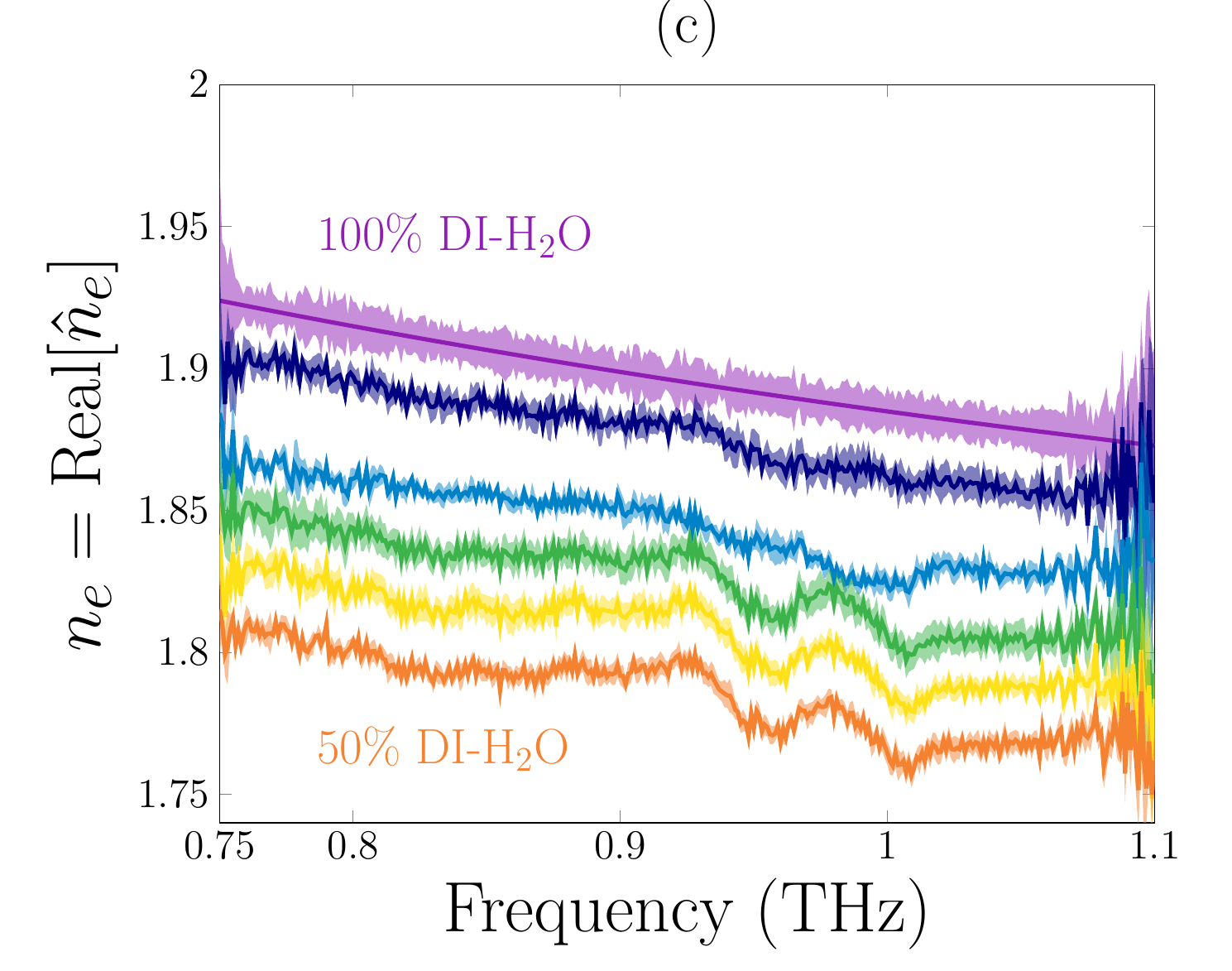}
\includegraphics[width=0.49\linewidth, clip=true, trim=6mm 0mm 7mm 0mm]{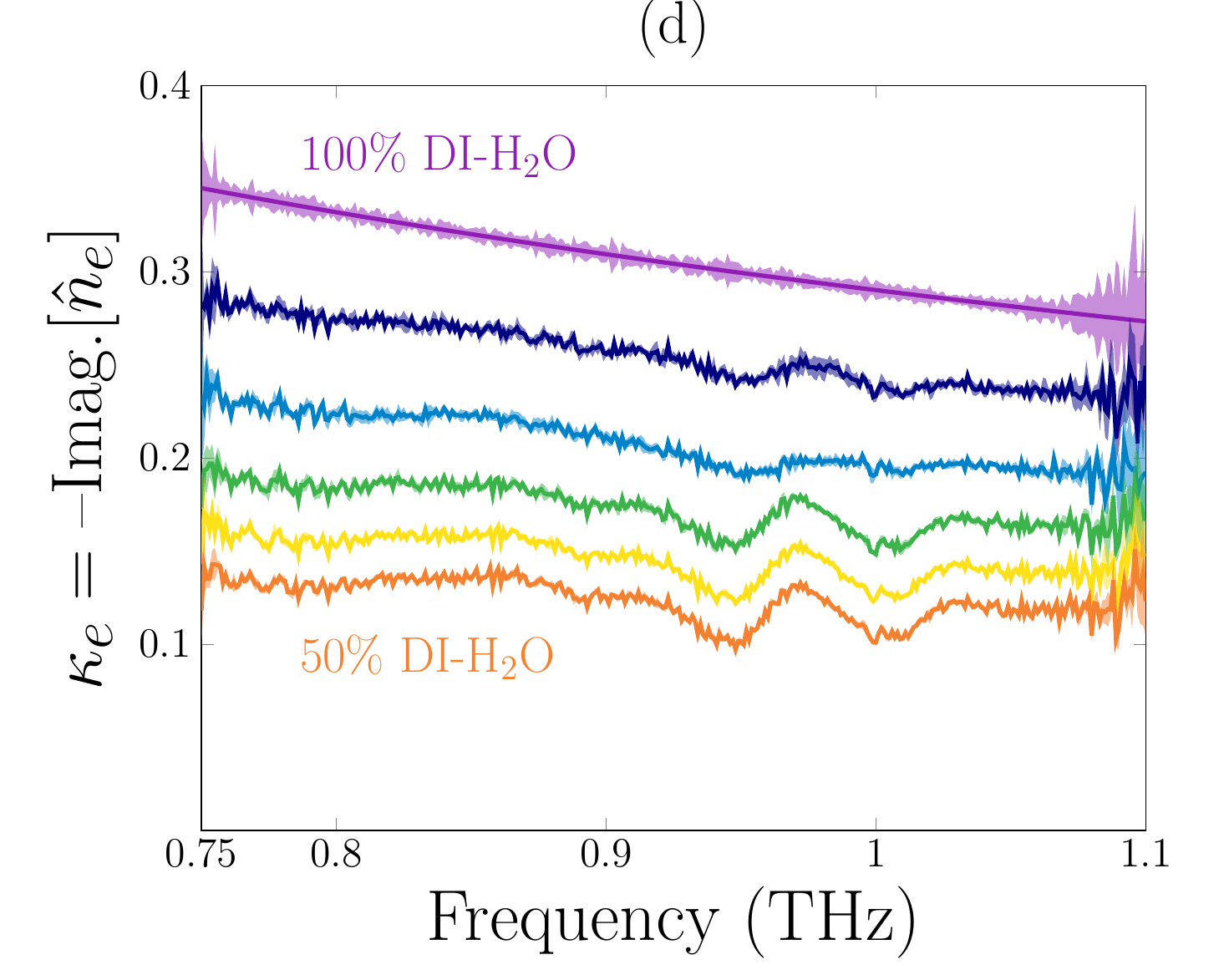}
\caption{Transmission measurements can resolve magnitude and phase differences between high-loss aqueous samples. Transmission (a)~phase difference,  and (b)~power ratio with respect to reference measurement for the IPA/DI-$\ce{H_2O}$ samples (from 50\% to 100\% DI-$\ce{H_2O}$ in steps of 10\%). (c)~Real and (d)~imaginary part of effective refractive indices of the CPW with samples. Lines indicate the mean of five measurements, whereas shadows indicate the standard deviation.}
\label{fig:eff_RI}
\end{figure}

We measured the transmission of a 1~mm long CPW with the PDMS microfluidic channel containing IPA/DI water solutions from $\SI{0.75}{THz}$ to $\SI{1.1}{THz}$. For DI-water, the typical insertion loss was in the order of $\SI{25}{dB}$.
Fig.~\ref{fig:eff_RI}a-b shows the phase and magnitude, respectively, of the transmission for each sample normalized to the reference measurement (DI-water).
Solid lines represent the average of five successive measurements, whereas the shadows represent their standard deviation.
Both the normalized magnitude and phase consistently decrease for increasing IPA concentration, as expected from literature values \cite{Yomogida2010, Kindt1996}.
The artifacts observed around $\SI{0.95}{THz}$ and $\SI{1}{THz}$, which appear as resonances in both magnitude and phase of the transmission, were also observed without any microfluidic channel. A possible explanation is parasitic probe-to-probe coupling effects or calibration artifacts \cite{Phung2019}.
The measurements' noise increases noticeably after $\SI{1.05}{THz}$ due to a drop in the dynamic range at the end of the frequency extender band.
The relative error shows to be smaller for $\kappa_e$ than for $n_e$, implying that in this case, smaller sample changes can be detected with attenuation measurements than with phase measurements. For water measurements the SNR was typically higher than \SI{25}{dB}.

\begin{figure}[!h]
\centering
\includegraphics[width=0.48\textwidth, clip=true, trim=10mm 0mm 7mm 0mm]{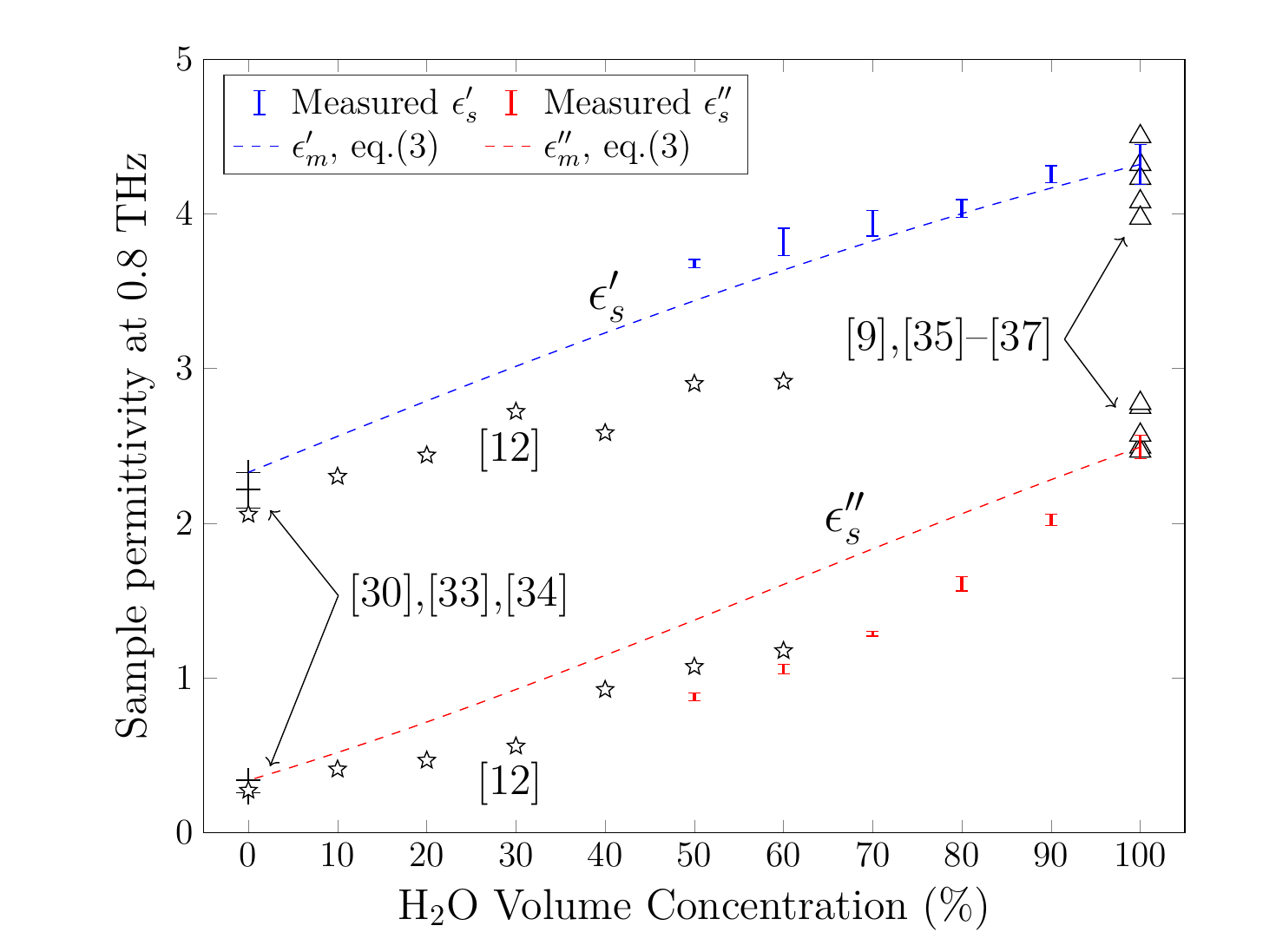}
\caption{Obtained sample permittivities show consistent change with varying water concentration. Permittivity of IPA/DI-$\ce{H_2O}$ samples vs. water concentration at $\SI{0.8}{THz}$ for this work (blue and red errorbars for 95\% confidence interval) and literature values. Literature values of ($+$) IPA  \cite{Yomogida2010,Yomogida2010a,Swithenbank2016}, ($\vartriangle$) water\cite{Kindt1996,Jepsen2007,Segelstein1981,Vinh2015}, and ($\medwhitestar$) IPA/$\ce{H_2O}$ mixtures measured with a similar on-chip method \cite{Swithenbank2017} are plotted. The dashed lines show Bruggeman's model approximation  for the change in water concentration, based in eq.~(3).}
\label{fig:eps_vsConc}
\end{figure}

The real and imaginary effective refractive indices are shown in Fig.~\ref{fig:eff_RI}.c-d, where all high-loss samples have been successfully resolved.
A relatively large effective length $l_s=\SI{250}{\um}$ was used since part of the liquid was found to extend between the PDMS and CPW substrate.
Fig.~\ref{fig:eps_vsConc} shows the samples' extracted real and imaginary permittivities at $\SI{0.8}{THz}$ versus sample DI-water concentration, showing error bars with 95\% confidence interval.
Both the real and imaginary parts of the permittivities change consistently with the changes of IPA concentration.
The analytically extracted sample permittivities  are also plotted with the permittivity of the IPA/water mixtures using literature values, using a modified Bruggeman's approximation for binary mixtures of polar liquids~\cite{Puranik1994}, with equation:
\label{eq:waterIPA}
 \begin{equation}
 \left[\frac{\epsilon_m-\epsilon_{_{\text{H}_2\text{O}}}}{\epsilon_{_{\text{IPA}}}-\epsilon_{_{\text{H}_2\text{O}}}}\right]\left[\frac{\epsilon_{_{\text{IPA}}}}{\epsilon_m}\right]^{1/3}=1-\left[a-(a-1)\alpha_{_{\text{H}_2\text{O}}}\right]\alpha_{_{\text{H}_2\text{O}}}
\end{equation}
where $\epsilon_{_{\text{H}_2\text{O}}}$ and $\epsilon_{_{\text{IPA}}}$ are literature values of the permittivities of water \cite{Kindt1996} and IPA \cite{Yomogida2010} respectively, $\alpha_{_{\text{H}_2\text{O}}}$ is the volume concentration of water, $\epsilon_{m}$ the permittivity of the mixture, and $a=1.33$ is the fitting factor for propanol-water samples \cite{Puranik1994}.

%% file: include/5_Conclusions.tex
\section{Conclusion}
This letter presents an on-chip dielectric spectroscopy method capable of measuring high-loss aqueous samples enabled by VNA and ground-signal-ground probes.
Similar attempts of broadband measurements of aqueous solutions with TDS had problems resolving the complex refractive index for high water content solutions at frequencies higher than $\SI{0.5}{THz}$ \cite{Swithenbank2017} or lost sensitivity from not placing sample where the sensor's field is strongest \cite{Kitagawa2006}.
The main reasons why the presented method could measure aqueous samples sensitively is due to (1) a higher coupling efficiency when exciting the CPWs using the contact probes compared to other methods which couple a free-space beam into the substrate via an antenna; (2) the higher average power of continuous-wave THz signals produced by the VNA compared to other pulsed methods, like TDS; (3) a low-loss design of the CPW. 
The high dynamic range (typically \SI{50}{dB} at calibration plane) allows to measure water with a sample length up to approximately \SI{0.6}{mm}.
Some drawbacks of this method are that (1) is limited to frequencies up to 1.1 THz, due to lack of contact probes at higher frequencies; (2) being limited to rectangular waveguide bands for a given setup, and (3) a higher cost.
In perspective, this method is a first step towards a miniaturized system for sensing high-loss aqueous samples at THz frequencies, which could provide an integrated, accurate, and controlled platform to study fundamental biological phenomena in their native environment.

%% file: include/6_Appendices.tex
%% file: include/7_Acknowledment.tex
\section*{Acknowledgment}

The authors would like to thank Mr. Mats Myremark for machining parts for the measurement setup and Dr. Kiryl Kustanovich for helping to fabricate the microfluidic channel. The devices were fabricated and measured in the Nanofabrication Laboratory and Kollberg Laboratory, respectively, at Chalmers University of Technology, Gothenburg, Sweden.

%% file: include/8_Biography.tex
\begin{IEEEbiography}[{\includegraphics[width=1in,height=1.25in,clip,keepaspectratio]{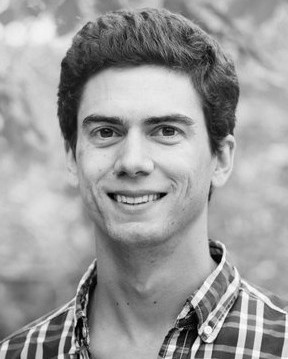}}]{Juan Cabello-S\'{a}nchez}
(S' 17) was born in Madrid in 1992. He received the Bachelor diploma in Electrical Engineering and Computer Science (ABET-accredited) from the Technical University of Madrid, Spain, in 2015, and MSc in Wireless, Photonics and Space Engineering from Chalmers University of Technology, in Gothenburg, Sweden, in 2017.

From July 2017 he is pursuing his PhD in the THz and Millimetre-wave Laboratory at Chalmers University of Technology.
\end{IEEEbiography}

\begin{IEEEbiography}[{\includegraphics[width=1in,height=1.25in,clip,keepaspectratio]{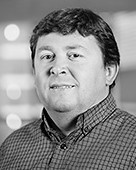}}]{Vladimir Drakinskiy}
Vladimir Drakinskiy was born in Kurganinsk, Russia, in 1977. He received the Diploma degree in physics and informatics (with honors) from the Armavir State Pedagogical Institute, Armavir, Russia, in 2000, and the Postgraduate degree from Moscow State Pedagogical University, Moscow, Russia, in 2003.

From 2000 to 2003, he was a Junior Research Assistant in the Physics Department, Moscow State Pedagogical University. Since 2003, he has been in the Department of Microtechnology and Nanoscience, Chalmers University of Technology, Gothenburg, Sweden. During 2003–2005, he was responsible for mixer chips fabrication for the Herschel Space Observatory. Since 2008, he has been a Research Engineer with the Department of Microtechnology and Nanoscience, Chalmers University of Technology. He is currently responsible for terahertz Schottky diodes process line at MC2, Chalmers University of Technology. His research interests include microfabrication and nanofabrication techniques, detectors for submillimeter and terahertz ranges, and superconducting thin films.
\end{IEEEbiography}

\begin{IEEEbiography}[{\includegraphics[width=1in,height=1.25in,clip,keepaspectratio]{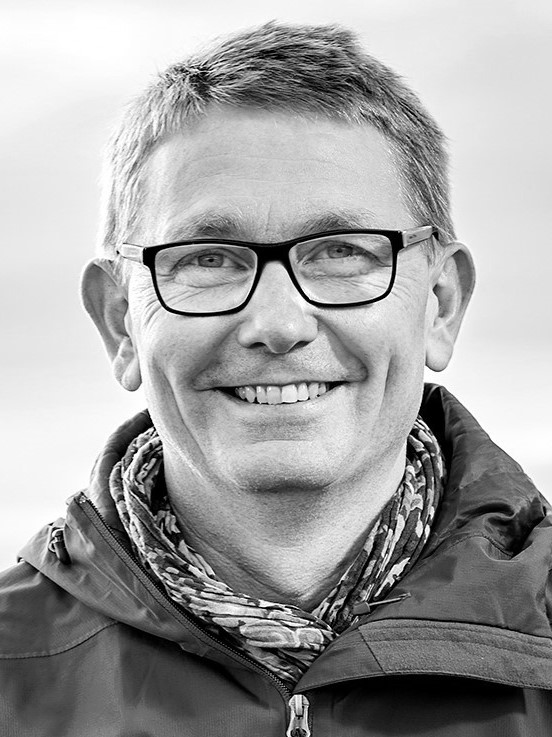}}]{Jan Stake}
(S' 95--M' 00--SM' 06) was born in Uddevalla, Sweden, in 1971. He received the M.Sc. degree in electrical engineering and the Ph.D. degree in microwave electronics from the Chalmers University of Technology, Goteborg, Sweden, in 1994 and 1999, respectively.

In 1997, he was a Research Assistant with the University of Virginia, Charlottesville, VA, USA. From 1999 to 2001, he was a Research Fellow with the Millimetre Wave Group, Rutherford Appleton Laboratory, Didcot, U.K. He was then a Senior RF/Microwave Engineer with Saab Combitech Systems AB, until 2003. From 2000 to 2006, he held different academic positions with the Chalmers University of Technology and, from 2003 to 2006, was also the Head of the Nanofabrication Laboratory, Department of Microtechnology and Nanoscience, MC2. During Summer 2007, he was a Visiting Professor with the Submillimeter Wave Advanced Technology Group, Caltech/JPL, Pasadena, CA, USA. He is currently a Professor and the Head of the Terahertz and Millimetre Wave Laboratory, Chalmers University of Technology. He is also a Cofounder of Wasa Millimeter Wave AB, Goteborg, Sweden. His research interests include graphene electronics, high frequency semiconductor devices, THz electronics, submillimeter wave measurement techniques (“THz metrology”), and THz in biology and medicine.

Prof. Stake served as Editor-in-Chief for the IEEE TRANSACTIONS ON TERAHERTZ SCIENCE AND TECHNOLOGY from 2016 to 2018.
\end{IEEEbiography}

\begin{IEEEbiography}[{\includegraphics[width=1in,height=1.25in,clip,keepaspectratio]{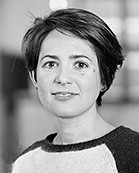}}]{Helena Rodilla}
(M' 16--SM' 20) was born in Salamanca, Spain, in 1982. She received the B.S. and Ph.D. degrees in Physics from the University of Salamanca, Salamanca, Spain, in 2006 and 2010, respectively.

From 206 to 2010, she was with the Electronics Group, Department of Applied Physics, University of Salamanca, Spain, where her research interest was semiconductor physics. From 2011 to 2013 she was Postdoctoral Researcher with the Microwave Electronics Laboratory, Department of Microtechnology and Nanoscience (MC2), Chalmers University of Technology, Gothenburg, Sweden, where she worked on very low-noise InP HEMTs for cryogenic low noise amplifiers. Since 2013 she has been with the Terahertz and Millimetre Wave Laboratory, MC2, Chalmers University of Technology, Gothenburg, Sweden, where she became Associate Professor in 2020. Her current research interests include the use of millimeter wave and terahertz technology in life science applications, sensing and on-wafer terahertz measurements.

\end{IEEEbiography}




